\documentstyle[psfig]{mn}
\begin{document}

\title[4U~1820-30 and Ser X-1]
{Radio detections of the neutron star X-ray binaries 4U~1820-30 and Ser X-1 in
soft X-ray states}
\author[S. Migliari et al.]
{S. Migliari$^1$\thanks{migliari@science.uva.nl},
R. P. Fender$^1$\thanks{rpf@science.uva.nl}, M. Rupen$^2$, S. Wachter$^3$,
P. G. Jonker$^4$, 
\newauthor  J. Homan$^5$, M. van der Klis$^1$\\ 
\\
$^1$ Astronomical Institute `Anton Pannekoek', University of Amsterdam, and
Center for High Energy Astrophysics, Kruislaan 403, \\ 1098 SJ, Amsterdam, The
Netherlands.\\ $^2$ National Radio Astronomy Observatory, Socorro, NM 87801,
USA\\ $^3$ SIRTF Science Center, California Institute of Technology, MS 220-6,
Pasadena, CA 91125\\ $^4$ Institute of Astronomy, Madingley Road, CB3 0HA,
Cambridge, UK \\ $^5$ MIT Center for Space Research, 77 Massachusetts Avenue,
Cambridge, MA 02139 \\ }

\maketitle

\begin{abstract}
We present the analysis of simultaneous X-ray (RXTE) and radio (VLA)
observations of two atoll-type neutron star X-ray binaries: 4U~1820--30 and
Ser X-1. Both sources were steadily in the soft (banana) X-ray state during
the observations. We have detected the radio counterpart of 4U~1820--30 at
4.86~GHz and 8.46~GHz at a flux density of $\sim0.1$~mJy. This radio source is
positionally coincident with the radio pulsar PSR~1820--30A. However, the
pulsar's radio emission falls rapidly with frequency ($\propto\nu^{-3}$) and
we argue that the X-ray binary's radio emission is dominant above
$\sim2$~GHz. Supporting this interpretation, comparison with previous
observations reveals variability at the higher radio frequencies that is
likely to be due to the X-ray binary. We have detected for the first time the
radio counterpart of Ser~X-1 at 8.46~GHz, also at a flux density of $\sim
0.1$~mJy. The position of the radio counterpart has allowed us to
unambiguously identify its optical counterpart. We briefly discuss
similarities and differences between the disc-jet coupling in neutron star and
black hole X-ray binaries. In particular, we draw attention to the fact that,
contrary to other states, neutron star X-ray binaries seem to be more radio
loud than persistent black hole candidates when the emission is `quenched' in
the soft state.

\end{abstract}


\section{Introduction}
Many works suggest that the radio emission in X-ray binaries, even when no
spatial structure is resolved, originates in jet-like outflows (Hjellming \&
Han 1995; Falcke \& Biermann 1996; Dhawan, Mirabel \& Rodr\'\i{}guez 2000;
Fomalont, Geldzahler \& Bradshaw 2001; Fender 2004). A X-ray/radio correlation
is therefore usually interpreted as a disc/jet coupling in the systems.  In
black hole candidate (BHC) X-ray binaries a connection between radio emission
and X-ray emission is already clear. Studies of BHCs in the low/hard state
showed a strong correlation between X-ray and radio fluxes over more than
three orders of magnitude in accretion rate (Hannikainen et al. 1998; Corbel
et al. 2000; Corbel et al. 2003; Gallo, Fender \& Pooley 2003). For X-ray
luminosities greater than a few per cent Eddington, where the X-ray spectra
soften dramatically, the radio emission seems to be `quenched' (Fender et
al. 1999; Corbel et al. 2001; Gallo et al. 2003).

Low magnetic field neutron star (NS) X-ray binaries have been classified,
based on X-ray spectral and timing properties, in two distinct classes whose
names recall the shape they trace in the colour-colour diagram (CD): Z-type
and atoll-type (Hasinger \& van der Klis 1989). The position of the source in
the CD defines its X-ray state. In atoll sources the hardest X-ray state is
called `island' state and the softest `banana' state (see Hasinger \& van der
Klis 1989 for details). Atoll-type X-ray binaries in the hard state share many
X-ray timing and spectral properties with BHCs in the low/hard state (e.g. van
der Klis 1994). However, our understanding of the relation between radio
activity and X-ray state in these NS sources remains sketchy. This is mainly
due to the fact that their radio luminosites are significantly less than in
BHs (Fender \& Hendry 2000), resulting in a rather small number of atoll
sources with identified radio counterparts (Hjellming \& Han 1995; Berendsen
et al. 2000).  Recently, Migliari et al. (2003) found the first evidence for a
disc-jet coupling in an atoll source, analysing simultaneous X-ray and radio
observations of the NS X-ray binary 4U~1728--34, when the source was in the
hard (i.e. mainly island) state. We found positive ranking correlations
between the radio flux density and X-ray flux and between the radio flux
density and X-ray timing features. We also confirmed quantitatively that in
the hard state black hole binaries are $\sim30$ times more 'radio loud' than
atoll-type NSs at a similar Eddington fraction (see also Fender \& Kuulkers
2001).

In this paper we analyse simultaneous X-ray and radio observations of two
atoll-type NS sources in the X-ray soft (i.e. middle-upper banana) state:
4U~1820--30 and Ser X-1 (4U~1837+04).

\subsection{4U 1820-30}

4U 1820--30 is a low-mass X-ray binary located in the globular cluster
NGC~6624 (Giacconi et al. 1974). A distance of $7.6\pm0.4$~kpc (Heasley et
al. 2000) has been derived from optical observations. Grindlay et al. (1976)
discovered thermonuclear X-ray bursts in the source, indicating a neutron star
as compact object. From X-ray burst properties a distance of $\sim 6.6$~kpc
can be derived (Vacca, Lewin \& van Paradijs 1986; see Kuulkers et al. 2003).
Smale, Zhang \& White (1997) reported the discovery of kHz QPOs in 4U
1820--30. A correlation between the frequency of the kHz QPOs and the position
in the CD was found by Bloser et al. (2000). In particular, in 4U 1820--30 the
kHz QPOs seem to be present while the source is in the lower banana state and
absent when in the upper banana. The broadband 2--50~keV spectrum of the
source is well fit with a typical model for atoll source in the soft state: a
blackbody (or a multitemperature disc blackbody) plus a cutoff power law or
with a Comptonisation model (e.g. {\em CompTT}: Titarchuck 1994) plus a
blackbody component (e.g. Bloser et al. 2000). The radio source at the
position of 4U 1820--30 was detected from observations with the Very Large
Array (VLA) at 1.4~GHz by Geldzahler (1983) with a flux density of
$2.44\pm0.37$~mJy and then by Grindlay \& Seaquist (1986) at the same
frequency with a flux density of $0.49\pm0.12$~mJy. The identification of this
radio source as the radio counterpart of the X-ray binary is
controversial. Johnston \& Kulkarni (1992) first identified the radio source
they found at 1.4~GHz at the position of the X-ray binary as the radio
counterpart of 4U~1820--30. Later, based on the steepness of the spectrum of
the radio source (a non-detection at 8.5~GHz with rms$\sim20$~$\mu$Jy; see
also Biggs et al. 1994), they (Johnston \& Kulkarni 1993) identified the radio
source as a pulsar. The radio position of this pulsar, PSR 1820--30A (Biggs et
al. 1990; Biggs et al. 1994; Stappers 1997) is coincident with the optical
counterpart of the X-ray binary (Sosin \& King 1995), and with the most recent
VLA detection at 1.5~GHz and 5~GHz (Fruchter
\& Goss 2000). These two sources are spatially unresolved. This makes the
identification of the real origin of the observed radio emission difficult.

\subsection{Ser X-1}
 
Ser X-1 was discovered in X-rays in 1965 (Friedmann, Byram \& Chubb 1967). The
source shows thermonuclear X-ray bursts, reported for the first time by Swank
et al. (1976). Ser X-1 is one of the three NS binary systems in which
simultaneous X-ray and optical bursts have been observed (Hackwell et
al. 1979).  Christian \& Swank (1997) derived a distance of $\sim8.4$~kpc from
burst properties.  The X-ray energy spectrum of the source is reasonably well
fit with the same model used for 4U 1820--30: a soft component, blackbody or
multicolor disc blackbody, and a Comptonised component (e.g. Oosterbroek et
al. 2001).  The optical counterpart was first identified as a blue star by
Davidsen (1975). Subsequently, it turned out that the optical counterpart was
in fact two unresolved stars (Thorstensen, Charles \& Bowyer 1980). The two
Davidsen stars are called DN (north) and DS (south), and DS (now called MM
Ser) is the proposed optical counterpart (Thorstensen et al. 1980). Wachter
(1997) found that DS is itself the superposition of two previously unresolved
stars, DSw (west) and DSe (east), separated by $1\arcsec$. The brightest of
the two stars (DSe) was suggested to be the real counterpart of Ser X-1 by
Wachter (1997). Hynes et al. (2003) deblended the spectra of the DSe and DSw
and found further spectral confirmation of Wachter (1997) suggestion. At radio
wavelengths Grindlay \& Seaquist (1986) reported upper limits of $<0.4$~mJy at
5~GHz.

\section{Observations and data analysis}

We have analysed simultaneous radio and X-ray observations of 4U 1820--30 and
Ser~X-1 performed with the VLA and with the Rossi X-ray Timing Explorer
(RXTE). The dates of the observations are shown in Table~1.

%#############TABELLA####################

\begin{table}%[!b]
\centering
\caption{Modified Julian Day (MJD), 2--10~keV unabsorbed flux (F$_{2-10 keV}$)
in erg s$^{-1}$ cm$^{-2}$, radio flux density at 4.86 GHz (F$_{4.86 GHz}$) and
at 8.46 GHz (F$_{8.46 GHz}$) in mJy for seven VLA observations of 4U 1820--30
(five of which simultaneous with RXTE) and for one VLA observation
(simultaneous with RXTE) of Ser X-1. Error bars are 1$\sigma$ errors.}
\label{tabspecB}
\vspace{0.2cm}
\begin{tabular}{l l l l}
\hline
\hline
MJD & F$_{2-10~keV}$          & F$_{4.86~GHz}$ & F$_{8.46~GHz}$\\
    & (erg s$^{-1}$ cm$^{-2}$) & (mJy)          & (mJy)         \\
\hline
\multicolumn{4}{c}{4U 1820-30} \\
\hline
52480$^{\rm a}$   & 7.9$\times10^{-9}$ & $0.16\pm0.10$  & $0.14\pm0.04$ \\
52482             & 8.7$\times10^{-9}$ & $0.13\pm0.10$  & $0.08\pm0.04$ \\
52484             & 8.7$\times10^{-9}$ & $0.15\pm0.09$  & $0.10\pm0.04$ \\
52487             & 9.0$\times10^{-9}$ & $0.17\pm0.09$  & $0.10\pm0.04$ \\
52489             & 9.2$\times10^{-9}$ & $0.08\pm0.09$  & $0.06\pm0.05$ \\
52492$^{\rm b,c}$ & \ldots             & $0.09\pm0.16$  & $0.04\pm0.06$ \\
52494$^{\rm b}$   & \ldots             & $-0.07\pm0.10$ & $0.09\pm0.04$ \\
Averaged              & 8.7$\times10^{-9}$ & $0.13\pm0.04$  & $0.10\pm0.02$ \\
\hline
\multicolumn{4}{c}{Ser X-1}\\
\hline
52421 & 4.4$\times10^{-9}$ & $0.05\pm0.04$ & $0.08\pm0.02$ \\
\hline
\end{tabular}
\flushleft
{\bf a:} the 2--10 keV unabsorbed flux is the averaged flux of the four RXTE
observations performed on MJD 52480; 
{\bf b:} not simultaneous with RXTE;
{\bf c:} bad weather
\end{table}
%#############################################

\subsection{VLA}

We have analysed seven observations of 4U~1820--30 (whose durations range from
about 40 min to two hours) and one observation (of about five hours) of Ser
X-1 with the VLA at 4.86 and 8.46~GHz. During the 4U~1820--30 observations the
VLA was in B configuration, and during the Ser~X-1 observation in A
configuration. For 4U~1820--30 we have used 1331+305 (3C286) as the flux
calibrator and 1820--254 (J2000 RA $18^{h}20^{m}57^{s}.8487$
Dec. $-25^{\circ}28^{\prime}12\arcsec.587$) as the phase calibrator. For
Ser~X-1 we have used both 0137+331 (3C48) and 1331+305 (3C286) as flux
calibrators and 1824+107 (J2000 RA $18^{h}24^{m}02^{s}.8554$
Dec. $-10^{\circ}44^{\prime}23\arcsec.772$) as the phase calibrator. Flux
densities of the two sources measured at both frequencies are shown in
Table~1. Much of the data of 4U 1820--30 on MJD 52492 had to be discarded due
to bad weather, leading to the larger errors in Table~1. The best-fit
coordinates, from fits in the image plane assuming a Gaussian model, for the
radio counterpart of 4U~1820--30 are J2000 RA $18^{h}23^{m}40^{s}.4820$
$\pm0^{s}.0088$ Dec. $-30^{\circ}21^{\prime}40\arcsec.12$ $\pm0\arcsec.16$,
and for the radio counterpart of Ser~X-1 are J2000 RA
$18^{h}39^{m}57^{s}.557$ $\pm0^{s}.010$
Dec. $+05^{\circ}02^{\prime}09\arcsec.50$ $\pm0\arcsec.14$. The
naturally-weighted synthesized beam width for the combined data of 4U~1820--30
is of $3.49\times1.46$ arcsec in position angle $-1.2\degr$ at 5~GHz and of
$1.96\times0.80$ arcsec in position angle $-2.2\degr$ at 8~GHz, and for
Ser~X-1 is of $1.51\times0.79$ arcsec in position angle $+66.2^{\circ}$ at
5~GHz and of $0.77\times0.46$ arcsec in position angle $+67.1^{\circ}$ at
8.3~GHz. There is no evidence for any spatial extension to the radio
counterparts of either source. Neither 4U~1820--30 nor Ser~X-1 shows
significant radio variability over the time of monitoring.

\subsection{RXTE}

%##########FIGURE##############################
\begin{figure}
\psfig{figure=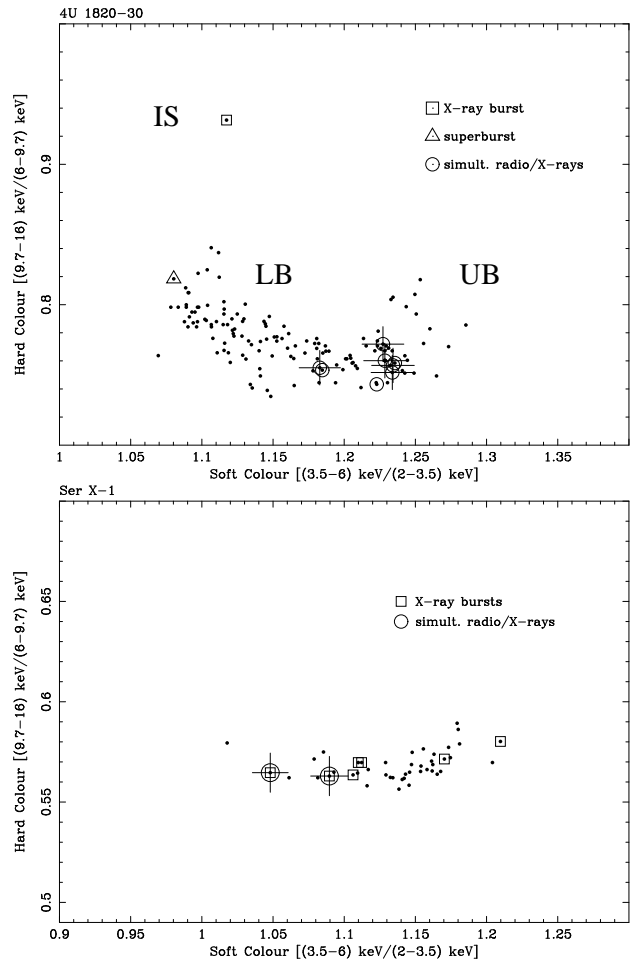,angle=0,width=8.2cm}\psfig{figure=./figure/serx1_CD.ps,angle=-90,width=8.2cm}
\caption{Mean colours (with PCU2 only) of each of all the observations of
4U~1820--30 (top panel) and Ser X-1 (bottom panel) in the RXTE public
archive. The observations marked with open circles are simultaneous with radio
(VLA) observations and those marked with open squares are the observations
that show X-ray bursts. In 4U 1820--30 (top panel) the open triangle shows the
observation in which the `superburst' was discovered (Strohmayer \& Brown
2000). Crosses indicate 90 per cent error bars.}
\label{CD_1820}
\end{figure}
%######################################################################

For the RXTE observations we have used data from the Proportional Counter
Array (PCA; for spectral and timing analysis) and the High Energy X-ray Timing
Experiment (HEXTE; only for spectral analysis). We have used the PCA {\tt
Standard2} data of the proportional counter unit 2 (PCU2; on in all the
observations) to produce the CD of all the RXTE observations of 4U~1820--30
and Ser X-1 available in the public archive. A soft colour and a hard colour
are defined as the count rate ratio (3.5--6)~keV/(2--3.5)~keV and
(9.7--16)~keV/(6--9.7)~keV, respectively. We have normalised the colours of
4U~1820--30 and Ser~X-1 to the colours of the Crab calculated with the closest
observation available to each observation analysed. Assuming the steadiness of
the Crab energy spectrum, this normalisation allows us to compare observations
in different epochs minimising shifts in the CD of instrumental origin. In
Fig.~\ref{CD_1820} we show the mean colours (black dots) of each of the public
RXTE observations of 4U~1820--30 (top panel) and Ser~X-1 (bottom panel). The
observations marked with open circles are simultaneous with radio (VLA)
observations. We have analysed in detail (X-ray spectral and timing analysis)
only these simultaneous radio/X-ray observations. The open squares mark the
observations in which X-ray bursts are observed in the PCA light curve. In the
CD of 4U~1820--30 (Fig.~\ref{CD_1820}, top) the open triangle shows the
observation in which the so-called `superburst' was discovered (Strohmayer \&
Brown 2000).

\subsubsection{Spectral analysis}

For the spectral analysis we have used PCA {\tt Standard2} and HEXTE {\tt
Standard Mode} data. For PCA data reduction we have subtracted the background
estimated using {\tt pcabckest} v3.0, produced the detector response matrix
with {\tt pcarsp} v8.0, and analysed the energy spectrum in the range
3--20~keV. For HEXTE data we extracted energy spectra (channels 15--61) from
cluster A, subtracted the background, corrected for deadtime using the
standard FTOOLS V5.2 procedures and analysed the spectra between 20 and 50
keV. A conservative systematic error of 0.75\% was added to the PCA data to
account for uncertainties in the calibration.

In the case of 4U~1820--30, all the 3--50~keV energy spectra obtained
simultaneously with radio observations are well fit with a CompTT model
(Titarchuk 1994) corrected for photoelectric absorption (without any need for
an extra blackbody components; see Bloser et al. 2000); in the worst case we
obtain $\chi^{2}_{re}=1.1$ with 49~$d.o.f.$ ($\chi^{2}_{re}=\chi^{2}/d.o.f.$).
The equivalent hydrogen column density N$_{\rm H}$ is fixed to
$3\times10^{21}$~cm$^{-2}$ (e.g. Bloser et al. 2000). Using this model we
obtain (Wien) temperatures kT$_{0}$ of the seed photons around 0.6~keV and
temperatures of the plasma of about 2.5--3~keV, with a plasma optical depth
$\tau \sim 7$.

For Ser X-1, using the same model that successfully fit the spectra of 4U
1820--30, i.e. CompTT corrected for photoelectric absorption (N$_{\rm H}$ was
fixed to 5$\times10^{21}$~cm$^{-2}$; e.g. Oosterbroek et al. 2001), we cannot
obtain a $\chi^{2}_{re}$ less than 1.8 (50~$d.o.f.$). The energy spectrum
shows an excess between 6--7~keV. Therefore we added a Gaussian emission line
around 6.5~keV to the model obtaining a better fit with $\chi^{2}_{re}=0.8$
(47~$d.o.f$). The CompTT model fit gives parameters values similar to those
found for 4U~1820--30 (i.e. kT$_{0}\sim0.7$~keV, temperatures of the plasma of
about 2.6~keV and $\tau\sim 6$). In Table~1 we show the unabsorbed 2--10~keV
fluxes of Ser~X-1 and of 4U~1820--30 using the CompTT model (plus a Gaussian
emission line in the case of Ser X-1).

\subsubsection{Timing analysis}

For the production of the power spectra we have used {\tt event} data with a
time resolution of 125 $\mu$s. We rebinned the data in time to obtain a Nyquist
frequency of 4096~Hz. For each observation we created power spectra from
segments of 128s length, using Fast Fourier Transform techniques (van der
Klis 1989 and references therein); we removed detector drop-outs from the
data, but no background subtraction was performed. No deadtime corrections
were done before creating the power spectra. We averaged the power spectra and
subtracted the Poisson noise spectrum applying the method of Zhang et
al. (1995), shifted in power to match the spectrum between 3000 and 4000~Hz
(see Klein-Wolt, Homan \& van der Klis in prep.). The Leahy normalisation was
applied (Leahy et al. 1983) and then we converted the power spectra to squared
fractional rms. We have fitted the power spectra with a multi-Lorentzian model
(e.g. Belloni, Psaltis \& van der Klis 2002 and references therein), and
plotted in the $\nu P_{\nu}$ representation (with $P_{\nu}$ the normalized
power and $\nu$ the frequency).
%##########FIGURE############################## 
\begin{figure}
\centering
\psfig{figure=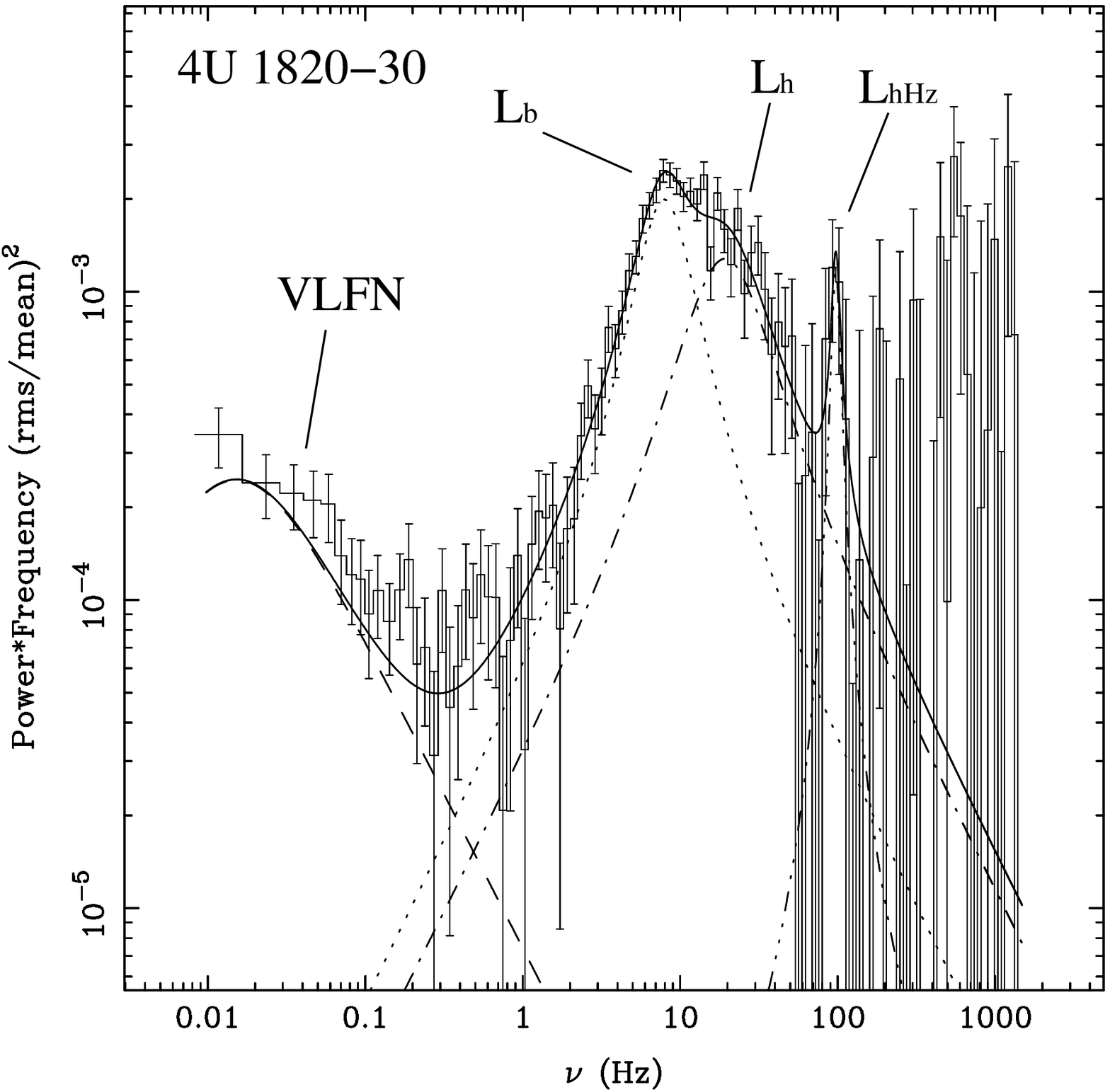,angle=0,width=8.2cm}\psfig{figure=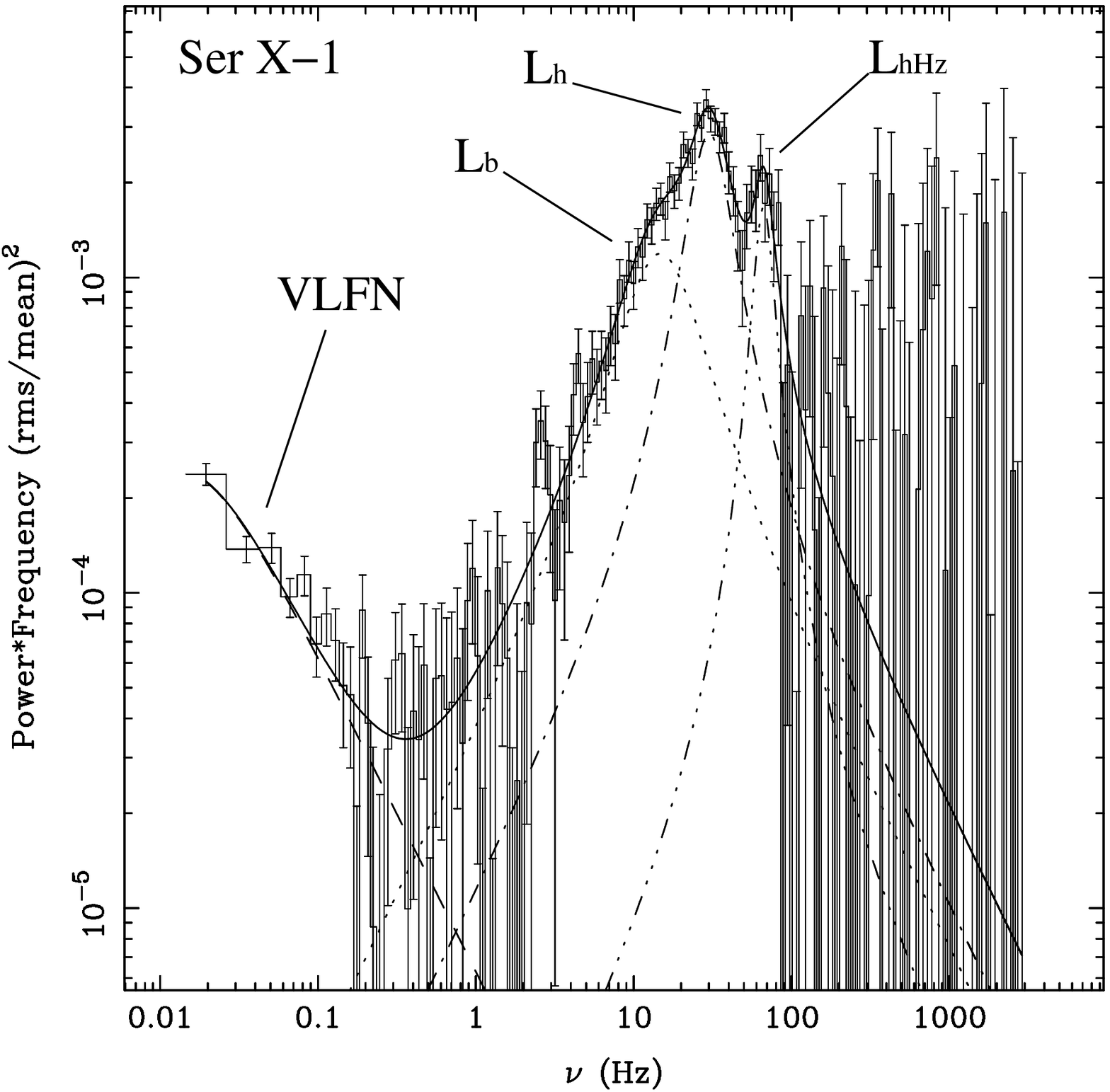,angle=0,width=8.2cm}
\caption{Power spectra of 4U~1820--30 on MJD 52482 (top panel) and of the
second observation on MJD 52421 of Ser X-1 (bottom panel) with the best-fit
model. Indicated are the four Lorentzian components fitting the very low
frequency noise (VLFN), the break frequency (L$_{b}$), the high frequency
feature (L$_{h}$) and the hecto Hz QPO (L$_{hHz}$) (but see \S~2.2.2).}  
\label{pow}
\end{figure}
%######################################################################
We need two to four Lorentzians to fit the seven power spectra of 4U~1820--30:
a broad Lorentzian to fit the very low frequency noise (VLFN), one or two
Lorentzians around 10--20~Hz to fit the break of the peaked noise component
(L$_{b}$; see van Straaten, van der Klis \& M\'endez 2003 for details on
terminology) and a higher frequency feature (L$_{h}$). In one observation (MJD
52482, see Fig.~\ref{pow}) we also need one narrow Lorentzian to fit a QPO
around 100~Hz (L$_{hHz}$).
We have fit the two power spectra of Ser X-1 with four Lorentzians: a broad
Lorentzian for the VLFN, one at 15~Hz, one around 30~Hz and, only in one of
the two observations (see Fig.~\ref{pow}) a narrow one at $\sim 70$~Hz. The
last three features are consistent with being identified either as L$_{b}$,
L$_{h}$ and L$_{hHz}$ (as in the 4U~1820--30 power spectrum), or L$_{b2}$,
L$_{b}$ and L$_{h}$.

\section{Results and discussion}

\subsection{4U 1820--30}

%##########FIGURE############################## 
\begin{figure}
\centering
\psfig{figure=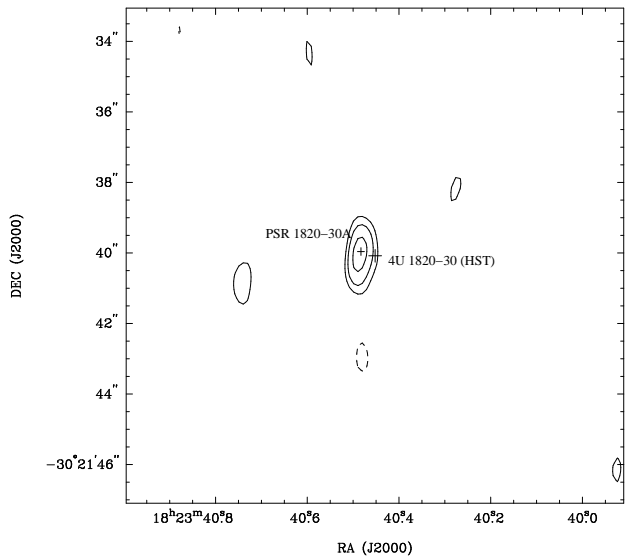,angle=0,width=8.2cm}
\caption{Naturally-weighted VLA radio contour plot of the averaged data of
4U~1820--30 at 8.46~GHz. We have chosen the contour interval as the rms of the
image (0.015~mJy) and plot contours at -4, -2.83, 2.828, 4, 5.657 times the
contour interval. The naturally-weighted synthetized beam width is of
$1.96\times0.80$ arcsec in position angle $-2.2^{\circ}$. The crosses indicate
the Hubble Space Telescope (HST) position of the X-ray binary and the radio
timing position of the pulsar PSR~1820--30A (see Table~2 and \S~3.1). Sizes of
crosses indicate $1\sigma$ errors on the position.
}
\label{map_1820}
\end{figure}
%######################################################################

The CD of 4U 1820--30 (Fig.~\ref{CD_1820}, top) shows that the simultaneous
radio/X-ray observations are in the middle-upper banana. This is consistent
with the power spectra characteristics: the presence of a VLFN (the integrated
fractional rms of the fitting Lorentzian is $\sim3$\%) is typical of atoll
sources in a soft X-ray state (i.e. middle-upper banana: see e.g. van Straaten
et al. 2002); the absence of kHz QPOs characterises the power spectra of
4U~1820--30 in the middle-upper banana (Zhang et al. 1998; Bloser et
al. 2000). The source is detected at 8.46~GHz with a flux density of
$0.10\pm0.02$~mJy and marginally detected ($\sim3\sigma$) at 4.86~GHz with a
flux density of $0.13\pm0.04$ in the combined data (see Table~1). The mean
(dual-frequency) radio spectrum has a spectral index $\alpha=-0.48\pm0.62$
(where $S_{\nu}\propto\nu^{\alpha}$ and $S_{\nu}$ is the radio flux density at
a frequency $\nu$) consistent with either an optically thin or an optically
thick spectrum. Biggs et al. (1994) reported radio detections (at 0.4~GHz,
0.6~GHz, 1.4~GHz and 1.7~GHz) of a pulsar, PSR 1820-30A. The position of this
pulsar (see Stappers 1997 for a more accurate timing position) is coincident
within $2\sigma$ with the optical position of the X-ray binary 4U~1820--30
(Sosin \& King 1995), and with the position of our radio detection (see
Table~2 and Fig.~\ref{map_1820}). From the positions listed in Table~2 and
using a distance to the sources of $\sim7$~kpc, we infer a lower limit on the
physical separation between PSR 1820-30A and 4U~1820--30 of the order of few
$10^{16}$~cm. 

In Fig.~\ref{map_1820} we directly compare the positions of the
sources plotting the positions listed in Table~2 on the VLA radio map. To do
so, we have to take into account errors in the VLA coordinates due to the
phase transfer. Therefore we have to increase the error bars when plotting
other positions atop the VLA map: we added (in quadrature) to the non-VLA
(i.e. the X-ray binary optical and the pulsar) positions an additional error
of 0.098~arcsec.  Moreover, in the case of the PSR~1820-30A (timing) position,
there is a systematic offset due to the transfer from the timing frame to the
VLA frame. Based on Fomalont et al. (1992), we add (in quadrature) an error of
0.005~arcsec in right ascension and 0.02~arcsec in declination to the pulsar
position errors. Two other pulsars are identified in NGC 6624: PSR~1820--30B
(Biggs et al. 1990; Biggs et al. 1994), which is too far from the 4U~1820--30
optical position to contaminate our radio detection, and PSR 1820--30C
(Chandler 2002), for which no coordinates are reported. In Fig.~\ref{map_1820}
we show the VLA radio contour map of 4U~1820--30 at 8.46~GHz. The crosses
indicate the Hubble Space Telescope (HST) optical position of the X-ray binary
(Sosin \& King 1995) and the Parkes radio timing position of the pulsar
PSR~1820--30A (Stappers 1997).

\begin{figure}
\centering
\psfig{figure=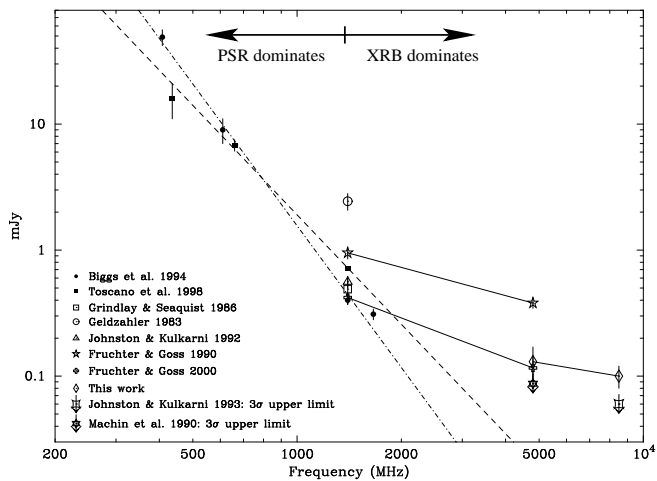,angle=0,width=8.6cm}
\caption{Radio flux density vs frequency of the published radio
observations of the pulsar PSR~1820--30A and the X-ray binary 4U~1820--30. The
dashed line is the best-fit power law of the pulsar observations from Toscano
et al. (1998; black dots) and the dashed-dotted line is the best-fit power law
of the pulsar observations from Biggs et al. (1994; black squares). The solid
lines are the power law fits of the dual-frequency radio observations at
$1.5$~GHz and $5$~GHz (open stars: Fruchter \& Goss 1990; open
crosses: Fruchter \& Goss 2000) and at $5$~GHz and $8.5$~GHz (open
diamonds: this work).} 
\label{1820_spec} 
\end{figure}

How do we know whether our radio detection comes from the X-ray binary or from
the pulsar PSR 1820--30A, since they are positionally concident?
Fig.~\ref{1820_spec} shows the broadband ($0.4-1.5$~GHz) spectrum of the radio
source.  The dashed-dotted line is the best-fit power law of the data from
Biggs et al. (1994) and the dashed line is the best-fit power law of the data
from Toscano et al. (1998); both of these fits are for the low-frequency
($<2$~GHz) radio spectrum and are typical for radio pulsars. The radio
spectrum of Toscano et al. (1998; the flatter of the two spectra in
Fig.~\ref{1820_spec}) is very steep, with $\alpha\sim-2.9$. The radio flux
density at 1.4~GHz is $\sim0.7$~mJy. This spectrum would predict
$\sim4$~$\mu$Jy at 8.5~GHz, thirty times less than we detect (open diamonds in
Fig.~\ref{1820_spec}). The spectrum of the pulsar from Biggs et al. (1994),
with $\alpha\sim3.7$, is even more discrepant. The overall spectrum we observe
in Fig.~\ref{1820_spec} seems to be the superposition of two main spectra,
coming from two physically distinct but spatially unresolved sources. We
suggest that below $\sim2$~GHz the pulsar dominates the radio emission with a
steep spectrum, while the X-ray binary dominates the radio emission above
$\sim 2$~GHz with a flatter spectrum. Other radio detections above
$\sim1.5$~GHz show large variability of the flux: at 4.8~GHz the flux density
varies from a non-detection with an rms of 0.020~mJy (Johnston \& Kulkarni
1993) to a detection of $0.38\pm0.035$~mJy (Fruchter \& Goss 1990). Previous
dual-frequency radio detections at 1.5~GHz and 5~GHz (Fruchter \& Goss 1990;
Fruchter \& Goss 2000; solid lines in Fig.~\ref{1820_spec}) show flat spectra
consistent with the spectrum we find with our detections at 5~GHz and
8.5~GHz. This further supports the idea that the X-ray binary dominates above
$\sim1.5$~GHz and therefore suggests that flux variations in this frequency
range are due to 4U~1820--30. The possibility that the flattening in the radio
spectrum at high frequency is due to emission from a pulsar wind nebula (PWN)
seems unlikely. First of all, the source shows a large radio flux variability
(more than one order of magnitute at 1.4~GHz; see Fig.~\ref{1820_spec}),
unlike PWN. Moreover, either a relatively dense interstellar medium (ISM)
and/or a high space velocity are required for the pulsar to power a wind
nebula. The former condition makes globular clusters, lacking in dense ISM, an
unlikely environment in which to find PWN (no PWN in globular clusters are, in
fact, known). In the case of a high space velocity, the pulsar would have
already escaped the central region of the cluster by now. The consistency with
the Ser~X-1 radio detection (e.g. similar radio luminosity at a similar X-ray
luminosity and state) further supports our interpretation (see below).

%###########TABLE#############################################
 \begin{table}%[!b]
\centering
\caption{Positions (with $1\sigma$ errors) of the optical and
radio counterparts of the X-ray binary 4U~1820--30 and the radio (timing)
position of the pulsar PSR~1820--30A.} 
\label{tabspecB}
\vspace{0.2cm}
\begin{tabular}{l l l l}
\hline
\hline
  & Pos. &J2000 & errors$^{a}$\\
\hline
VLA source$^{b}$  & RA & $18^{h}23^{m}40^{s}.4820$ & $\pm0^{s}.0088$\\
                         & Dec &$-30^{\circ}21^{\prime}40\arcsec.12$&$\pm0\arcsec.16$  \\         
4U~1820--30 (HST)$^{c}$  & RA &  $18^{h}23^{m}40^{s}.453$&$\pm0^{s}.012$  \\
                         & Dec& $-30^{\circ}21^{\prime}40\arcsec.08$   &
$\pm0\arcsec.15$  \\ 
PSR~1820--30A$^{d}$    & RA &  $18^{h}23^{m}40^{s}.4840$ &$\pm0^{s}.0006$ \\
                         & Dec& $-30^{\circ}21^{\prime}39\arcsec.96$   &
$\pm0\arcsec.05$  \\  

\hline

\end{tabular}
\flushleft
{\bf a:} To directly compare the positions on the VLA map, as we do in
Fig.~\ref{map_1820}, we have to add (in quadrature) an error of 0$\arcsec$.098
to the non-VLA positions, plus an additional error to the pulsar position due
to a systematic offset between the timing and the VLA frame (see \S~3.1).
{\bf b:} Our detection
{\bf c:} Sosin \& King 1995
{\bf d:} Stappers 1997
\end{table}
%#############################################

%##########FIGURE############################## 
%######################################################################

\subsection{Ser X-1}

\begin{figure}
\centering
\psfig{figure=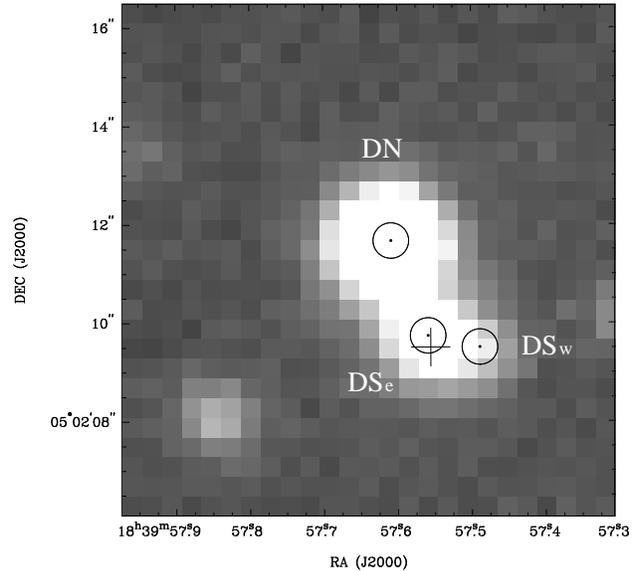,angle=0,width=8.2cm}
\caption{Portion of the optical ($R$-band) image of the field of MM Ser, with
the three stars: DN, DSe and DSw (see Wachter 1997). The position (with
$1\sigma$ errors) of the three stars are marked with open circles; the
position (with $1\sigma$ errors) of the radio counterpart is indicated with a
cross (see Table~3 and \S~3.2).}
\label{opt}
\end{figure}

The CD of Ser~X-1 (Fig.~\ref{CD_1820}, bottom) indicates that the source is in
the middle banana X-ray state during the radio observations. This is supported
by the variability properties, e.g. the presence of a VLFN (rms$\sim3$\%),
which are typical for an atoll source in its soft state (e.g. van Straaten et
al. 2002). In both 4U~1820--30 and Ser~X-1 we find similar spectral components
in the energy spectra, which confirms that the sources are both in a similar
(soft) state. We have detected for the first time the radio counterpart of Ser
X-1, with a flux density at 8.46~GHz of $0.08\pm0.02$~mJy (see Table~1). This
detection is consistent with the previous upper limit of $<0.4$~mJy at 5~GHz
of Grindlay \& Seaquist (1986). Wachter (1997) showed that MM Ser, the optical
counterpart of Ser~X-1, is actually the superposition of two stars separated
by $1\arcsec$. Therefore the position of the radio counterpart allows us to
identify the correct optical counterpart. In Fig.~\ref{opt} we show the
optical image of the three stars (DN, DSe and DSw) with positions and errors
marked with open circles. The optical image is a 400 sec R band exposure
obtained with the CTIO 0.9m telescope on 1996 July 11 (for details on the
reduction and calibration of the image see Wachter 1997). An astrometric
solution for the $3^{\prime}\times3^{\prime}$ frame was derived utilizing 40
stars from the USNO-A2.0 catalog and the IRAF task ``ccmap''. The errors on
the optical positions are a combination of the uncertainties due to the USNO
A-2.0 system ($0\arcsec.3$) and to the transfer of that coordinate system onto
the Ser~X-1 frame ($0\arcsec.2$). The cross indicates our VLA position (with
$1\sigma$ errors) of the radio counterpart. We clearly see that the radio
position is coincident within uncertainties with DSe. From the spectral
identification of Hynes et al. (2003) and our positional identification we can
definitely confirm that DSe is MM Ser, the actual optical counterpart of Ser
X-1.

%###########TABLE#############################################
\begin{table}
\centering
\caption{Optical positions (with $1\sigma$ errors) of the DN, DSe
and DSw stars (open circles) and the position (with $1\sigma$ errors) of the
radio counterpart (cross) of the X-ray binary Ser~X-1.}
\label{tabspecB}
\vspace{0.2cm}
\begin{tabular}{l l l l}
\hline
\hline
  & Pos. &J2000 & errors$^{a}$\\
\hline
Ser X-1 (VLA)&RA & $18^{h}39^{m}57^{s}.557$ & $\pm0^{s}.010$\\
            &Dec & $+05^{\circ}02^{\prime}09\arcsec.50$ &$\pm0\arcsec.14$  \\
DN        &RA & $18^{h}39^{m}57^{s}.61$              & $\pm0^{s}.02$  \\ 
          &Dec & $+05^{\circ}02^{\prime}11\arcsec.67$ & $\pm0\arcsec.36$  \\ 
DSe       &RA & $18^{h}39^{m}57^{s}.56$              & $\pm0^{s}.02$  \\
          &Dec & $+05^{\circ}02^{\prime}09\arcsec.74$ & $\pm0\arcsec.36$  \\
DSw       &RA & $18^{h}39^{m}57^{s}.49$             & $\pm0^{s}.02$  \\ 
          &Dec & $+05^{\circ}02^{\prime}09\arcsec.51$ & $\pm0\arcsec.36$  \\
\hline
\end{tabular}
\flushleft
{\bf a:} To directly compare the positions on the optical image, as we do in
Fig.~\ref{opt}, we have to add (in quadrature) an error of 0$\arcsec$.36 to
the VLA position of Ser X-1 (see \S~3.2) 
\end{table}
%#############################################

\subsection{Radio:X-ray correlation in X-ray binaries}

The simultaneous radio and X-ray detections of 4U~1820--30 and Ser~X-1 adds
new information to the radio/X-ray relations in X-ray binaries. These two
atoll sources are both steadily in the soft (banana) state. The radio emission
of BHCs in the low/hard state increases as the X-ray flux increases and then
decreases drastically (it is `quenched') by a factor of $>50$, when the source
enters a softer (high/soft or intermediate) X-ray state. In 4U 1728--34 in the
hard state (mostly island with two excursions to the lower banana) a similar
correlation has been observed (Migliari et al. 2003). One data point (the one
with the highest X-ray luminosity) lay off the correlation, possibly (even if,
oddly, this observation is still in the island state) indicating a radio
`quenching'. We have now detected the radio counterparts of two NS systems in
the soft state (middle-upper banana, with X-ray luminosities higher than those
of 4U~1728--34) at radio luminosities close to that found in 4U~1728--34 for
its highest X-ray luminosity observation. However, in the case of these NS
systems the reduction in radio luminosities between the brightest hard state
and the soft state seems to be only of a factor of $\sim10$. This indicates
that {\em if} there is a radio `quenching' also in atoll sources, this would
be less extreme than in BHCs. Furthermore, if we compare the luminosities we
have observed for these NS sources and the upper limits on the radio
luminosities of the BHCs in the soft state (quenching), we note that in BHs
the radio luminosity is lower than in NSs. In fact, we have detected
4U~1820--30 and Ser~X-1 at 8.5~GHz with a radio luminosity of
$\sim5.0\times10^{28}$ erg s$^{-1}$ (using a distance of 7~kpc) and
$\sim5.8\times10^{28}$ erg s$^{-1}$ (using a distance of 8.4~kpc),
respectively. The BHC XTE~1550-564 when `quenched' has a radio luminosity
upper limit of $<1.8\times10^{28}$ erg s$^{-1}$ (measured at 8.5~GHz and using
a distance to the source of 6~kpc; Corbel et al. 2002) and GX~339-4 of
$<2.6\times10^{28}$ erg s$^{-1}$ (measured at 5~GHz and using a distance of 6
kpc; Fender et al. 1999). Therefore, NS X-ray binaries in the soft state (when
the jet, at least for BHCs, seems to be suppressed) are more luminous in the
radio band than BH X-ray binaries. This is contrary to other X-ray states, in
which the BHCs are more radio loud, and seems to indicate that a 'residual'
radio emission is present in the NS systems. This difference with BHCs might
be related to the presence of a solid surface and/or to the NS magnetic
field. Recent theoretical works (e.g. Meier 2001) seems to support the idea
that an efficient jet-like outflow production is associated with a
geometrically thick accretion disc (i.e. low/hard state in BHCs), while a
suppression of the (observable) jet should happen when the disc become
geometrically thin (i.e. high/soft state in BHCs). Speculating within this
picture, assuming the same jet production processes in NS and BH X-ray
binaries (see Fender et al., 2004) a possible explanation might be that the
magnetic field of the NS could interact with the accretion disc, which in the
softer states should be closer to the compact object, keeping the disc thick
enough not to suppress completely the jet production mechanisms.

\section{Conclusions}

We have detected two atoll-type neutron star X-ray binaries, 4U 1820--30 and
Ser X-1, at radio wavelengths. In particular:

\begin{itemize}

\item We argue that we have detected the radio counterpart of 4U~1820--30,
which can be spectrally distinguished from the positionally coincident radio
pulsar PSR 1820--30A. The radio emission of the X-ray binary is dominant at
higher frequencies, above $\sim 2$~GHz and shows a flatter spectrum.

\item We have detected for the first time the radio counterpart of Ser X-1;
this radio detection allows us to identify star DSe as the optical counterpart
of Ser~X-1.

\item Both 4U~1820--30 and Ser~X-1 are detected at radio wavelengths while
steadily in the banana (soft) state. This is different from 4U~1728--34 which
has been detected at radio wavelengths in the banana state during transient
excursions from the island.

\item The radio luminosities in the soft state of 4U~1820--30 and Ser~X-1 are 
higher than the radio luminosities in the soft state of BHCs (for which
the jet is drastically suppressed). If the pattern of the radio flux vs. X-ray
flux of atoll sources is analogous to BHCs, then radio `quenching' in soft
state of atoll sources does not seem to be as extreme as in BHCs.

\end{itemize}

%##########FIGURE############################## 
%######################################################################

\section*{Acknowledgements}
SM and RPF would like to thank Ben Stappers for very useful discussions. The
National Radio Astronomy Observatory is a facility of the National Science
Foundation operated under cooperative agreement by Associated Universities,
Inc.

\end{document}